\documentstyle[12pt]{article}
\newlength{\extraspace}
\newlength{\extraspaces}
\newcommand{\newsection}[1]{
\vspace{15mm}
\pagebreak[3]
\addtocounter{section}{1}
\setcounter{subsection}{0}
\setcounter{footnote}{0}
\begin{flushleft}
{\large\bf \thesection. #1}
\end{flushleft}
\nopagebreak
\medskip
\nopagebreak}

\begin{document}
\addtolength{\baselineskip}{.7mm}

\thispagestyle{empty}

\begin{flushright}
{\sc PUPT}-1423\\
September 1993
\end{flushright}
\vspace{.3cm}

\begin{center}
{\Large\bf{An Extension of Birkhoff's Theorem to
a Class \\[2mm] of 2-d Gravity Theories Containing Black
Holes}}\\[20mm]
{\sc Youngjai Kiem}\\[3mm]
{\it Joseph Henry Laboratories\\[2mm]
 Princeton University\\[2mm]
Princeton, NJ 08544\\[2mm]
E-mail: ykiem@puhep1.princeton.edu}
\\[30mm]

{\sc Abstract}
\end{center}

     A class of 2-dimensional models including 2-d dilaton gravity,
spherically symmetric reduction of $d$-dimensional
Einstein gravity  and other related theories are
classically analyzed.  The general analytic solutions in the absence
of matter fields other than a U(1) gauge field are obtained under a
new gauge choice and recast in the conventional conformal gauge.
These solutions imply that Birkhoff's theorem, originally applied
to spherically symmetric 4-d Einstein gravity, can be applied to
all models we consider.  Some issues related to the coupling of
massless scalar fields and the quantization are briefly discussed.

\noindent

\vfill

\newpage

\newsection{Introduction}

     The quantum physics of black holes has always been an exciting problem
with a lot of controversies.  In recent years,  some of the major
intricacies related to the information loss problem via the
Hawking radiation process have been addressed within the framework
of two-dimensional dilaton gravity model, proposed by Callan, Giddings, Harvey
and Strominger (CGHS) \cite{CGHS}. This model is believed to capture many
essential features of the conventional four dimensional black hole theory
and, at the same time, is analytically manageable to  allow more
detailed calculations.  Dilaton gravity is a special case
of a more general family of two-dimensional models that in particular
includes the spherically symmetric reduction of Einstein gravity.
{}From a phenomenological point of view, it is therefore important to
understand to what extent the results about dilaton gravity that have
been obtained so far are independent of the specific choice of
parameters.

     The purpose of this work is to understand the relation
between the different two-dimensional models at the
classical level.  We find that they can be given a unified
treatment classically and, in the absence of matter fields
other than a U(1) gauge field, all of them satisfies
Birkhoff's theorem, originally applied to spherically
symmetric Einstein gravity.  This theorem is a direct manifestation
of the fact that there are no propagating degrees of freedom
in these theories.  We establish this result
by getting the general solutions of the equations
of motion under a particular choice of gauge.  This gauge,
generalized from an approach found in
Ref.\cite{hawking}, enables one to obtain
explicit general solutions.  Following the detailed analysis of
these solutions, we will
discuss the issues of introducing matter fields and the
quantization of models we consider.

\newsection{The Extension of Birkhoff's Theorem}

The model we consider is given by the following action;
\begin{equation}
     I = \int d^2 x \sqrt{-g} e^{-2 \phi} [
      R + \gamma g^{\alpha \beta} \partial_{\alpha} \phi
      \partial_{\beta} \phi + V(e^{- \phi} )
       - \frac{1}{4} e^{\epsilon \phi} F^2 ] .
\label{oaction}
\end{equation}
$R$ denotes the scalar curvature and $F$, the curvature 2-form for an
Abelian gauge field.
The field  $\phi$ is the dilaton field and the parameters $\gamma$
and $\epsilon$ above are assumed to be arbitrary real parameters.
$V(e^{ - \phi} )$ is a generic real function of the dilaton field.
The action introduced above can be considered as a 2-d target space
effective action resulting from string theory. \cite{witten}
In view of this aspect, the loop corrections from string theory
can give non-trivial contribution to $V$.
Eq.(\ref{oaction}) also contains many models of interest as its special
cases.  For example, setting $\gamma = 2$ and $V(x)  =
constant/ x^2 $  gives the
spherically symmetric reduction of 4-d Einstein gravity
coupled with electromagnetic fields.  \cite{lowe}
As is well known, the static solution of this case is
given by Reissner-Nordstr\" om blackholes and the dilaton field
can be related to the usual radial coordinate of the metric.
Another important example is $\gamma = 4$ and $V = constant$
case that corresponds to two-dimensional dilaton gravity discussed in
Ref.\cite{CGHS}.
As can be seen in the calculations in this section, the
interpolating theories connecting
these two important theories continuously show very similar
behavior as far as the classical analysis is concerned.
A model proposed by Teitelboim \cite{teitelboim} can also be
described by the action, if we set $\gamma = 0$ and $V = constant$.

     The static solutions of Eq.(\ref{oaction}) were obtained in
Ref.\cite{nappi}.
When it comes to spherically symmetric four dimensional gravity in the absence
of matter fields, we have more general results, namely Birkhoff's
theorem. This theorem essentially asserts that the general solutions
of this problem are just static Schwarzschild solution in each local
coordinate patch. In this note we wish to extend this result to all
models described by Eq.(\ref{oaction}). To this end
we have to solve the equations of motion
\begin{equation}
D_{\alpha}D_{\beta} \Omega - g_{\alpha \beta} D \cdot D \Omega
 + \frac{\gamma}{8} (g_{\alpha \beta} \frac{(D \Omega )^2 }
{\Omega} - 2 \frac{D_{\alpha } \Omega D_{\beta} \Omega}
{\Omega} ) + \frac{1}{2} g_{\alpha \beta} \Omega V(\Omega )
\label{eomg}
\end{equation}
$$
 - \frac{1}{8} (g_{\alpha \beta} F^2 -  4 g^{\mu \nu}
F_{\alpha \mu} F_{\beta \nu} ) \Omega^{1 - \epsilon /2}
= 0
$$
\begin{equation}
R + \frac{\gamma}{4} ( \frac{(D \Omega )^2}{\Omega^2} - 2
\frac{D \cdot D \Omega}{\Omega} ) + \frac{d}{d \Omega}
(V( \Omega ) \Omega) - \frac{1}{4} (1- \frac{\epsilon}{2}
) \Omega^{- \epsilon /2} F^2 = 0 ,
\label{eomo}
\end{equation}
where we define $ \Omega = \exp (-2 \phi ) $
and $D$ denotes the covariant derivative.  The first of
the above equations is obtained by varying the action with respect
to the metric tensor and the second, with respect to the
dilaton field.

     The well known conformal gauge is not so convenient for
the classical analysis we are interested in.  Instead we choose
a different gauge where the metric tensor is of the form
\begin{equation}
g_{\alpha \beta} = \left[\begin{array}{cc}
                    - \alpha^2 & 0 \\
                    0          & \beta^2   \end{array} \right] .
\end{equation}
Furthermore, we choose coordinates in such a way that $x^1
\equiv  r  \equiv \exp (- \phi ) $ and require
$ [ \partial_0 , \partial_1 ] = 0$.  This procedure defines
a coordinate system up to the diffeomorphisms of the time
coordinate.  The scalar curvature in this coordinate system
has the following form;
\begin{equation}
    \sqrt{-g} R = 2 [
\partial_0 (\frac{\partial_0 \beta}{\alpha} )
 - \partial_1 (\frac{\partial_1 \alpha}{\beta} ) ] .
\end{equation}
The resulting Christoffel
symbols are calculated to be
\begin{equation}
     \Gamma_{11}^0 = \frac{\beta \partial_0 \beta}{\alpha^2},
  \ \   \Gamma_{01}^0 = \frac{\partial_1 \alpha}{\alpha},
  \ \   \Gamma_{00}^0 = \frac{\partial_0 \alpha}{\alpha}
\label{chris}
\end{equation}
$$
     \Gamma_{11}^1 = \frac{\partial_1 \beta}{\beta},
  \ \   \Gamma_{01}^1 = \frac{\partial_0 \beta}{\beta},
  \ \   \Gamma_{00}^1 = \frac{\alpha \partial_1 \alpha}{\beta^2} ,
$$
which is the standard result for a diagonal metric.
We note that in case of the 4-d spherically
symmetric Einstein gravity $x^1$ reduces to the usual radial coordinate
of the Schwarzschild geometry.

     Modulo total derivatives, the action in this gauge
is simplified  to yield
\begin{equation}
     I = \int d^2 x [ 4 r \frac{\partial_1 \alpha}{\beta}
+ \gamma \frac{\alpha}{\beta} + r^2 V(r) \alpha \beta
+ r^{2 - \epsilon } \frac{f^2}{2 \alpha \beta} ] ,
\label{raction}
\end{equation}
where the function $f \equiv \partial_0 A_r - \partial_r A_0$
and satisfies $F^2 = - \frac{2}{\alpha^2 \beta^2} f^2$.
We immediately see that the dynamics of the original action
looks greatly simplified in this gauge.  First of all, the
derivatives with respect to $x^0$ that, in spherically
symmetric 4-d gravity, corresponds to the time derivatives
appear only as total derivatives and, consequently, can be
completely thrown away.  Secondly, since there is only one
linear first-order $r$-derivative, the
resulting equations of motion in this gauge are first order
differential equations, not the generic second order
differential equations.  The equations of motions from
the action (\ref{raction}) are
\begin{equation}
     4r \frac{\partial_1 \alpha}{\beta^2} + \gamma
\frac{\alpha}{\beta^2} - r^2 V(r) \alpha + r^{2 - \epsilon}
 \frac{f^2}{2 \alpha \beta^2}    = 0
\label{reom0}
\end{equation}
\begin{equation}
     4r \frac{\partial_1 \beta}{\beta^2} + (\gamma - 4)
\frac{1}{\beta} + r^2 V(r) \beta - r^{2 - \epsilon}
\frac{f^2}{ 2 \alpha^2 \beta} = 0 ,
\label{reom1}
\end{equation}
along with the equations for gauge fields,
\begin{equation}
     \partial_0 ( \frac{r^{2- \epsilon }}{\alpha \beta} f ) = 0
\end{equation}
\begin{equation}
     \partial_1 ( \frac{r^{2- \epsilon}}{\alpha \beta} f ) = 0 .
\end{equation}
The equations for the abelian gauge field can be solved to
give
\begin{equation}
     f = \alpha \beta r ^{-2 + \epsilon} f_0 ,
\label{fsolution}
\end{equation}
where $f_0$ is a constant.
The trivial nature of the solution is understandable since the
purely radial motion of the system can not generate physical
(transversal) polarization states of photons.

    The gauge constraints resulting from the choice of our gauge
follow from Eqs.(\ref{eomg}) and (\ref{eomo}). They are
\begin{equation}
\frac{1}{\sqrt{-g}} \frac{\delta I}{\delta \Omega} = 0
\label{con1}
\end{equation}
and
\begin{equation}
\frac{1}{\sqrt{-g}} \frac{\delta I}{\delta g_{01}} = 0 ,
\label{con2}
\end{equation}
where $I$ represents the original action (\ref{oaction}).
Using Christoffel symbols (\ref{chris}) to explicitly write down covariant
derivatives, we obtain
\begin{equation}
\partial_0 \partial_1 \Omega - \frac{\partial_1 \alpha}{\alpha}
\partial_0 \Omega - \frac{\partial_0 \beta}{\beta}
\partial_1 \Omega - \frac{\gamma}{4} \partial_0 \Omega
\partial_1 \Omega = 0
\label{a1}
\end{equation}
from Eq.(\ref{con2}).  From Eq.(\ref{con1}),
we obtain
\begin{equation}
\frac{2}{\alpha \beta} (\partial_0 ( \frac{\partial_0 \beta}{\alpha} )
- \partial_1 (\frac{\partial_1 \alpha}{\beta} ) )
+ \frac{d}{d\Omega} (V(\Omega ) \Omega ) - \frac{1}{4}
(1- \frac{\epsilon}{2} ) \Omega^{- \epsilon /2} F^2
\label{a2}
\end{equation}
$$
 - \frac{\gamma}{4} [ \frac{1}{\Omega^2} (  \frac{ (\partial_0 \Omega
)^2}{\alpha^2} - \frac{ (\partial_1 \Omega )^2 }{\beta^2} )
$$
$$ + \frac{2}{\Omega} \{  - \frac{\partial_0^2 \Omega}{\alpha^2}
+ \frac{\partial_1^2 \Omega}{\beta^2}
 + \frac{1}{\alpha^2} ( \frac{\partial_0 \alpha}{\alpha} -
\frac{\partial_0 \beta}{\beta} ) \partial_0 \Omega
+ \frac{1}{\beta^2} ( \frac{\partial_1 \alpha}{\alpha}
- \frac{\partial_1 \beta}{\beta} ) \partial_1 \Omega \} ] = 0 .
$$
The crucial conditions from the definition of our coordinate
systems are $\partial_1 \Omega = 2r$ and $\partial_0 \Omega = 0$.
Plugging these conditions into Eq.(\ref{a1}) and Eq.({\ref{a2})
yields
\begin{equation}
2r \frac{\partial_0 \beta}{\beta} = 0
\label{a3}
\end{equation}
and
\begin{equation}
\frac{2}{\alpha \beta} ( \partial_0  ( \frac{\partial_0 \beta}
{\alpha} ) - \partial_1 ( \frac{\partial_1 \alpha}{\beta } ) )
- \frac{\gamma}{\Omega^{1/2} \beta^2} (\frac{\partial_1 \alpha}
{\alpha} - \frac{\partial_1 \beta}{\beta} )
\label{a4}
\end{equation}
$$
 + \frac{d}{d\Omega} (V(\Omega ) \Omega ) - \frac{1}{4}
(1- \frac{\epsilon}{2} ) \Omega^{\- \epsilon /2} F^2 = 0 .
$$
Since we are engaged in classical analysis,
Eq.(\ref{a4}) can be further simplified using the equations of motion
(\ref{reom0}) and (\ref{reom1}) along with the
classical solution of $F$ to remove $r$-derivatives of
$\alpha$ and $\beta$.  The result of this removal is
\begin{equation}
\frac{1}{\alpha \beta} \partial_0 (
\frac{\beta}{\alpha} (\frac{\partial_0 \beta}
{\beta} ) )  = 0 .
\label{a5}
\end{equation}
We now find that the constraint reduces to
\begin{equation}
 \partial_0 \beta = 0
\label{con}
\end{equation}
from Eq.(\ref{a3}) and this automatically implies that
Eq.(\ref{a5}) is satisfied.

     Eqs.(\ref{reom0}) and (\ref{reom1}) can now be solved under
the simple constraint (\ref{con}).  The general solutions are
\begin{equation}
      \beta^2 = \frac{2 r^{2- \gamma /2}}
{C + \int_{r_0}^{r} x^{(6- \gamma )/2} (V(x) -
 f_0^2 x^{\epsilon -4} )  dx }
\label{rsol1}
\end{equation}
\begin{equation}
     \alpha^2 = T^2 (x^0 ) \frac{r^{- \gamma /2}}{2}
( C + \int_{r_0}^{r} x^{(6 - \gamma )/2} ( V(x) -
 f_0^2 x^{\epsilon -4} )  dx ) ,
\label{rsol0}
\end{equation}
where $T(x^0 )$ is an arbitrary function depending only on
$x^0$. $C$ and $r_0$ are arbitrary constants of integration.
The presence of the arbitrary function $T$ originates from the
arbitrariness involved in our definition of $(x^0 , r )$
coordinates, namely, the possible diffeomorphisms of the time coordinate.
This arbitrariness can be fixed by absorbing $T$ into $x^0$ by
properly redefining it.
We note that the constant $C$ could have involved
truly non-trivial $x^0$ dependence
if there were no constraint (\ref{con}).

     We can derive the static solutions of Eq.(\ref{oaction})
assuming all the relevant fields involved depends only on a single
variable, say $r$.  The results of this calculation are the same as
Eqs.(\ref{rsol1}) and (\ref{rsol0}) with the additional fact that
$T$ is strictly a constant.  Thus, we have proven a general result;
the general classical solutions of the action (\ref{oaction})
are same as the static solutions of the same action in each
properly defined local coordinate patch.  In other words, the
classical dynamics of the gravity models coupled with an Abelian
gauge field in the absence of other matter fields is locally
frozen.

\newsection{Aspects of the Analytical Solutions}

     To better understand the connection between the results
in the gauge of the previous section and the results in the
conformal gauge, it is desirable to recast our solutions
in conformal gauge. In terms of conformal coordinates $(x^+ ,
x^- )$, the metric should be written as
\begin{equation}
 ds^2 = - \alpha^2 (dx^0 )^2 + \beta^2 (dr)^2
= -e^{\rho} dx^+ dx^- .
\label{dconf}
\end{equation}
This condition is equivalent to
two sets of two partial differential equations
\begin{equation}
 \beta \partial_+ r = \pm \alpha \partial_+ x^0
\end{equation}
\begin{equation}
 \beta \partial_- r = \mp \alpha \partial_- x^0 ,
\end{equation}
along with an equation for the conformal factor $\rho$
\begin{equation}
e^{\rho} = - 4 \beta^2 \partial_+ r \partial_- r .
\end{equation}
We require $\partial_+ r \partial_- r < 0$ to fix the
orientation and make a choice of upper
signs.  The form of solutions (\ref{rsol1}) and (\ref{rsol0})
shows that the factors containing the coordinate $r$ and
the factors containing $x^0$ are simply multiplied.
Therefore, by a proper field redefinition, we can reduce the above
equations into the Laplacian equations in flat space for two
redefined fields.  The general solutions of
the PDEs obtained in this way are
\begin{equation}
\int_{r_1}^r \frac{2rdr}{C+ \int_{r_0}^{r}
x^{(6-\gamma )/2} (V(x) - f_0^2 x^{\epsilon - 4} ) dx }
  = X^+ (x^+ ) - X^- (x^- )
\label{csol0}
\end{equation}
\begin{equation}
\int_{t_0}^{x^0} T(t) dt = X^+ (x^+ ) + X^- (x^- ) ,
\end{equation}
where $r_1$ and $t_0$ are constants of integration and
$X^{\pm}$ are arbitrary chiral fields depending only on
$x^{\pm}$, respectively.  Using these solutions, the conformal
factor can be calculated to yield
\begin{equation}
e^{\rho} = 2 r^{ -\gamma / 2} [ C +
\int_{r_0}^{r} x^{ (6- \gamma )/2} ( V(x)
- f_0^2 x^{\epsilon -4 } ) dx ] \partial_+ X^+ \partial_- X^- .
\label{csol1}
\end{equation}
Combined with $r = e^{- \phi}$, Eqs.(\ref{csol0}) and
(\ref{csol1}) implicitly determine $\rho$ and $\phi$
via one left-moving and one right-moving field.
Particularly, in the case of 2-d dilaton gravity, these relations get
considerably simplified.  If we set $\gamma = 4$,
$V(r) = 4\lambda^2$, $C = -2 \lambda^2 M$, $f_0 = 0$
and conformally
transform $\exp (\pm 2 \lambda^2 X^{\pm})  \rightarrow
\pm X^{\pm}$, then the above solutions become
\begin{equation}
e^{- 2 \phi} = M - \lambda^2 X^+ X^-
\end{equation}
\begin{equation}
e^{\rho} = e^{2 \phi} \partial_+ X^+ \partial_- X^- ,
\end{equation}
which are the familiar linear dilaton vacuum
solution of CGHS model, where $M$ is the mass of the resulting
black hole. \cite{wadia}  In retrospect, this shows why the choice of
conformal gauge renders an analytically tractable approach in
2-d dilaton gravity, while this kind of approach is more
difficult in other cases, the complication being the difficulty of
the identification of chiral fields.

     In the spherically symmetric reduction of
4-d Einstein gravity case, i.e.,
$\gamma = 2$, $V(x) = 2/x^2$, $\epsilon = 0$, and $C = -4M$,
our solutions (\ref{rsol1}) and (\ref{rsol0}) reduce to
\begin{equation}
\alpha^2 = T^2 (x^0 ) ( 1 - \frac{2M}{r} + \frac{f_0^2 /2}{r^2} )
\end{equation}
\begin{equation}
\beta^2 = ( 1 - \frac{2M}{r} + \frac{f_0^2 /2 }{r^2} )^{-1} ,
\end{equation}
which represents the Reissner-Nordstr\" om black hole with mass $M$ and
electric charge $f_0/ \sqrt{2}$.
The $d$-dimensional Einstein gravity with the symmetry
group $SO(d-1)$ that is the $d$-dimensional generalization of the 4-d
spherically symmetric case can also be described by our action.
After the integration over angular coordinates and
the rescaling of the dilaton field which can be related to the
radial coordinate, we find that the effective action
for  these cases corresponds to Eq.(\ref{oaction}) with
$\gamma = 4 \frac{d-3}{d-2}$ and $V(x) = \mu (d) / x^{\frac{4}{d-2}}$.
Here, $d$ is the dimension of the space-time and $\mu (d)$ is a number
depending on $d$.  If we set $f_0 =0$, for simplicity, we get the
following general solutions;
\begin{equation}
\alpha^2 =  T^2 (x^0 )\frac{C + \frac{d-2}{2(d-3)} \mu (d)
r^{ 2 \frac{d-3}{d-2} } }{2 r^{2 \frac{d-3}{d-2}}}
\label{generald}
\end{equation}
\begin{equation}
\beta^2 = \frac{2 r^{\frac{2}{d-2}}}{ C + \frac{d-2}{2(d-3)}
\mu (d) r^{2 \frac{d-3}{d-2}}} .
\end{equation}
If $d=4$, this solution becomes the Schwarzschild metric.

     An interesting observation
is that the 2-d dilaton gravity case is same as
the $d = \infty$ limit of $d$-dimensional spherically symmetric
Einstein gravity.  In this limit, we have $\gamma \rightarrow 4$
along with $V(x) \rightarrow \mu (\infty ) = constant$ that
we can set $\mu (\infty ) = 4 \lambda^2$.  Taking $C = -2 \lambda^2 M$
as in the previous consideration
of the 2-d dilaton gravity, we can write the 2-d dilaton black hole metric
with mass $M$
in a form similar to the 4-d Schwarzschild metric,
\begin{equation}
ds^2 = - \lambda^2 ( 1- \frac{M}{r^2} ) T^2 (x^0 ) (dx^0 )^2
 + \frac{1}{1- \frac{M}{r^2} } \frac{dr^2}{ \lambda^2 r^2} ,
\label{dsch}
\end{equation}
which was obtained from Eq.(\ref{generald}) by taking the $d = \infty$
limit. At the level of the classical analysis, this shows
that CGHS model can be identified with the leading order theory of
the spherically symmetric reduction
of the finite dimensional Einstein gravity
in a $1/d$ expansion.  In analogy to the Schwarzschild metric, 2-d
dilaton black hole metric can also be maximally extended to form
a space-time similar to that of the maximally extended Kruskal
space-time.  In the limit of the strong coupling regime where
$r = e^{- \phi} \rightarrow 0$, i.e., $\phi \rightarrow + \infty$,
there exists a curvature singularity as can be explicitly seen
from Eq.(\ref{dsch}).

\newsection{Discussion}

     The analysis in this work provided us with the complete
classical solutions for many 2 dimensional models containing
black holes.  The questions to pursue from now on should be
at least two-fold; the coupling of matter fields other than
a U(1) gauge field should be considered to study the truly dynamical
formulation and Hawking evaporation of black holes.  The quantization
of the action considered here is another issue.

     Birkhoff's theorem, extended in this work, asserts there
is no truly dynamical evolution of the system described by
Eq.(\ref{oaction}).  Consequently, we can say that each and every
distinct classical solution of the action represents either
a distinct black hole state or a
vacuum state.  Considering the no-hair theorem that is valid
classically, we see that any additional matter coupling not shown in
Eq.(\ref{oaction}) should produce dynamic formation of black hole(s)
from incoming matter fields and
subsequent scattering into black hole(s) and/or outgoing matter fields.
To get a proper understanding of this complicated process, it is
imperative to consider additional matter couplings.
The gauge choice made in this note seems advantageous even
when we include non-trivial matter fields, e.g., massless
scalar fields, at least classically.  To be specific, we can
add to the original action
\begin{equation}
- \frac{1}{2} \int d^2 x \sqrt{-g} e^{-2\delta \phi} g^{\alpha \beta}
\partial_{\alpha} f \partial_{\beta} f ,
\end{equation}
the action for a massless scalar field where $\delta$ is a real
number.  The resulting equations of motion other than the one
for the scalar field are
\begin{equation}
\frac{\partial_1 \alpha}{\alpha} + \frac{\gamma}{8\Omega}
- \frac{\beta^2}{2} \Omega V(\Omega ) -
\frac{\Omega^{\delta}}{4} \{ \frac{\beta^2}{\alpha^2} (\partial_0
f )^2 + (\partial_1 f )^2 \} = 0
\end{equation}
\begin{equation}
\frac{\partial_1 \beta}{\beta} + \frac{\gamma}{8\Omega}
+ \frac{\beta^2}{2} \Omega V(\Omega ) -
\frac{\Omega^{\delta}}{4} \{ \frac{\beta^2}{\alpha^2} (\partial_0
f )^2 + (\partial_1 f )^2 \} = 0 ,
\end{equation}
where, for brevity, we did not include the U(1) gauge field and
used $(x^0 , \Omega ) = (x^0 , r^2 )$ coordinates. The main
virtue of this gauge choice is that the resulting gauge constraint
is very simple.  The gauge constraint (\ref{con2}) reduces to
\begin{equation}
\frac{\partial_0 \beta}{\beta} - \frac{1}{2} \Omega^{\delta}
\partial_0 f \partial_1 f = 0 ,
\label{dcon}
\end{equation}
which shows that the time variation in the mass of a black hole
results from the space and time variation of incoming and
outgoing scalar fields.  Just as in the case considered in
this work, the remaining gauge constraint (\ref{con1}) can
be written as
\begin{equation}
\frac{2}{\alpha \beta} \partial_0 \{ \frac{\beta}{\alpha}
( \frac{\partial_0 \beta}{\beta} - \frac{1}{2}
\Omega^{\delta} \partial_0 f \partial_1 f ) \} =0
\end{equation}
after somewhat lengthy calculations.
Thus, imposing Eq.(\ref{dcon}) automatically implies the other
constraint.  These relatively less complicated sets of first
order partial differential equations
may provide us with some further analytical understanding of
the classical picture of black hole formation, clarifying
the issue of gravitational back reaction.

     One benefit of our gauge choice other than the simplification
of classical analysis is the existence of a natural time-like
coordinate $x^0$.  This can be utilized to define a natural
Hamiltonian structure which, in principle, can be the basis of
canonical quantization.  The result of this quantization
will largely be topological in the
absence of additional matter fields, due to the lack of
local propagating degrees of freedom.
As to the quantization including additional
matter fields, we also note that
our solutions in conformal gauge can be very useful.  In the context
of two-dimensional dilaton gravity,  an explicit quantization of
the theory was given Ref.\cite{svv}. The fact that in this case the
most general form of the classical solutions are known was crucial
in their analysis.  The formal similarity of the class of solutions
described by Eqs.(\ref{csol1}) and
(\ref{csol0}) suggests the possibility that their analysis
can be generalized to some models described by our action. The
main interest,  from a phenomonological viewpoint, lies in the
quantization of spherically symmetric
4-d Einstein gravity. In this case there is a natural
(reflecting) boundary $r = 0$ where left-moving and right-moving
chiral fields can interact, while in case of dilaton gravity
this boundary had to be introduced by hand. This
distinction between the classical theories may lead to important
qualitative differences at the quantum level. We plan to address
the quantization of the models described by (\ref{oaction}) in a future
publication.

{\sc \bf Acknowledgement}

The author wishes to express his gratitude to H. Verlinde
for very useful discussions and reading of the text.  This work was
partially supported by NSF grant PHY-90-21984.

\end{document}